Exploring Neuronal Bistability at the Depolarization Block

Andrey Dovzhenok, Alexey S. Kuznetsov

Department of Mathematical Sciences and Center for Mathematical Biosciences, Indiana University Purdue University Indianapolis, Indianapolis, Indiana 46202, USA

**Corresponding author:**

Andrey Dovzhenok

Department of Mathematical Sciences

Indiana University Purdue University Indianapolis

402 N. Blackford St., Indianapolis, IN 46202, USA

E-mail: adovzhen@iupui.edu;   Phone: 317-278-6468;   Fax: 317-274-3460

**Abstract**

Many neurons display bistability – coexistence of two firing modes such as bursting and tonic spiking or tonic spiking and silence. Bistability has been proposed to endow neurons with richer forms of information processing in general and to be involved in short-term memory in particular by allowing a brief signal to elicit long-lasting changes in firing. In this paper, we focus on bistability that allows for a choice between tonic spiking and depolarization block in a wide range of the depolarization levels. We consider the spike-producing currents in two neurons, models of which differ by the parameter values. Our dopaminergic neuron model displays bistability in a wide range of applied currents at the depolarization block. The Hodgkin-Huxley model of the squid giant axon shows no bistability. We varied parameter values for the model to analyze transitions between the two parameter sets. We show that bistability primarily characterizes the inactivation of the $Na^+$ current. Our study suggests a connection between the amount of the $Na^+$ window current and the length of the bistability range. For the dopaminergic neuron we hypothesize that bistability can be linked to a prolonged action of antipsychotic drugs.

**Keywords**

Bistability; Hysteresis; Dopaminergic neuron; Hodgkin-Huxley model; excitability; spiking currents

**Introduction**

Bistability – coexistence of two firing modes in the same experimental conditions – has been documented in different types of neurons. Tonic spiking coexists with bursting [1] or with a different spiking mode [2] in leech heart cells. Bistability of bursting and spiking was also discovered in neuron R15 of the marine mollusk *Aplysia* [3]. In this paper, we focus on the bistability between a resting and tonic spiking states. This type of bistability was observed in different motor neurons [4-6]. The same type of bistability is hypothesized to be involved in short-term memory (discussed in [7]). In a bistable cell, a short signal triggers a long-lasting change in the firing, which encodes the last input. Altogether, bistability is common among neurons and endows them with richer forms of information processing.

In this study we focus on the bistability at the transition to the state called depolarization block – a silent state that occurs in every neuron when it receives excessive excitation. In vitro, a neuron enters depolarization block whenever the applied current exceeds a certain level. In a slightly different experiment, an iontophoresis current that supplies an excitatory neurotransmitter can also lead the neuron into depolarization block. Its minimal value that silences the neuron characterizes the neuron and the specific receptor (e.g. NMDA). Furthermore, depolarization block was suggested to explain the therapeutic action of antipsychotic drugs [8]. In schizophrenia and other diseases, the level of the neurotransmitter dopamine (DA) is abnormally high. Antipsychotics were shown to have a direct excitatory influence on the neurons releasing dopamine – dopaminergic neurons. This should further elevate the DA levels unless the DA neuron enters depolarization block and stops releasing dopamine. The effectiveness of the antipsychotics was linked to their ability to suppress DA neuron activity by depolarization block. DA neuron is one of two examples explored in this article.

Bistability at the transition to depolarization block has been observed in multiple neurons [4-7, 9]. However, it was not studied in models (but see [10]) and, more importantly, most experimental studies pay no attention to bistability at this transition. The terminology itself is not ready to account for two separate transitions – the stabilization of the silent state and the cessation of spiking. Which of these transitions should be called depolarization block? Which one is observed in experiments? This depends on the experimental protocol. What does it say about the neuron when the spiking and the silent states are both stable in a wide range of the applied current? We address these questions and prepare a theoretical basis for experimental studies of bistability. We investigate what factors contribute to bistability at the transition to depolarization block and propose to differentiate in experiments the two transitions involved.

Our results show that the silent state of depolarization block may be stable together with the tonic spiking state. In DA neurons, progressive depolarization block was proposed as a mechanism for the maximal therapeutic action of antipsychotic drugs [8]. Chronic administration of drugs used in treatment of schizophrenia results in silencing of the DA neurons due to depolarization block [18,19]. Taken together with our results, the DA neurons may stay in the silent state after lowering the dose or complete cessation of the drug administration because of the possible bistability between the silent and active states (Figure 1).

**Definitions**

We start by defining the notions of bistability and hysteresis in a dynamical system.

Definition 1. The lack of reversibility as a parameter is varied is called hysteresis [11].

Definition 2. A dynamical system having two coexisting attractors (stable solutions) is called bistable [12]. The solutions attract trajectories starting from different initial conditions and determine distinct long-term behavior.

Bistability is realized in a range of a parameter and is generally lost at bifurcations as a stable solution disappears or loses stability. Bistability for some range of the parameter is a necessary condition for hysteresis in any dynamical system.

The definition of depolarization block and the experimental protocols do not take into account bistability at a strong applied depolarization. The cessation of oscillations involves the loss of stability or the disappearance of the oscillatory solution and transition to the stable equilibrium state. This may occur by different scenarios.

Scenario 1: The oscillatory solution that corresponds to spiking decreases in amplitude to zero and merges with the equilibrium that corresponds to the silent state. The silent state becomes stable. This transition is a single supercritical Andronov-Hopf bifurcation, and it does not involve any hysteresis (see e.g. Figure 4C).

Scenario 2: The equilibrium state becomes stable by giving birth to an unstable oscillatory solution. This is a subcritical Andronov-Hopf bifurcation. Further increase in the applied current causes the unstable oscillatory solution to merge with the stable one, which corresponds to spiking (see e.g. Figure 4B). Both solutions disappear, and this transition is a saddle-node bifurcation of oscillatory solutions (limit cycles). This transition involves bistability because between the Andronov-Hopf and the saddle-node bifurcations, the stable equilibrium and the stable oscillatory solution coexist.

To correctly describe the range of applied current where both solutions are stable - bistability range, we introduce the following definitions:

Definition 3. We call the range of the applied current where the equilibrium state is unstable the instability range.

Definition 4. We call the range of the applied current where a stable oscillatory solution exists the oscillatory range.

When there is no bistability in the model, these parameter ranges coincide. On the other hand, when the oscillatory solution disappears in a saddle-node bifurcation of limit cycles, the oscillatory range extends to higher applied currents than the instability range. The difference between these ranges is exactly the bistability range.

The main objective of the current study is to reveal mechanisms that cause bistability in the spiking subsystem of a neuron comprised of the fast sodium and the rectifying potassium currents. Virtually every neuron has these currents and their interaction results in very different transitions between tonic spiking and the silent state. One example we consider in this article is the giant squid axon modeled by [13]. We call it the Hodgkin-Huxley (HH) neuron in the rest of the article. In the HH neuron, there is no bistability at the transition to depolarization block (Figure 4C). The transition from spiking to silence occurs through the supercritical Andronov-Hopf bifurcation. The other example we consider is the spiking subsystem of the dopaminergic (DA) neuron [14]. The DA neuron model displays strong bistability over a large range of applied depolarization. The model includes several currents that work in the voltage range below the spike initiation threshold, but bistability remains strong even when the model is reduced to spiking currents only. We call this reduced model the DA neuron in the rest of the paper for simplicity. After the reduction, the model includes exactly the same set of spiking currents and has the same structure and dimension as the model in [13]. Thus, it's not a different set of currents, but rather altered parameters of the same spike-producing currents that determine if the model displays bistability or not. In this paper, we identify particular parameters of the spiking currents that produce hysteresis and discuss physiological distinctions that characterize these currents in different neuron types.

**Conductance-Based Model**

Our DA neuron and the HH neuron are simply two different sets of parameters for the following conductance-based model. The model contains delayed rectifier potassium, fast sodium and leak currents and is given by the following system of differential equations:

$$C\frac{dv}{dt} = I_{app} - g_K n^4 (v - E_K) - g_{Na} m_\infty^3(v) h (v - E_{Na}) - g_L (v - E_L)$$

$$\frac{dn}{dt} = f_n \cdot (n_\infty(v) - n)/\tau_n(v)$$

$$\frac{dh}{dt} = f_h \cdot (h_\infty(v) - h)/\tau_h(v)$$

where $v$ is the membrane potential in mV, $n$ and $h$ are the activation gating variable for the $K^+$ current and the inactivation gating variable for the $Na^+$ current, correspondingly. The constant factors $f_n$ and $f_h$ in the equations above are originally set to unity and are only varied to study the effect of gating variables' kinetics (see Figure 7 below).

Thus the model is a 3-dimensional dynamical system with variables $v$, $n$ and $h$. Gating variables $m$, $n$ and $h$ have steady state voltage-dependent functions in the form $X_\infty(v) = 1/(1 + \exp(-(v - v_{Xh})/S_X))$ (where $X$ can be $m$, $n$ or $h$) and voltage-dependent time constant functions given by $\tau_X(v) = \tau_X^0 + \tau_X^1 \exp\left(-(v-\theta_X)^2/S_X^\tau\right)$ (where X is $n$ or $h$). The model parameter values for the HH neuron and the DA neuron are given in Table 1. Computer simulations were performed in XPPAUT [15] using the stiff method and a time step of 0.1 ms.

The comparison of the steady state functions of the DA neuron and the HH neuron shows that the half-activation/inactivation values for the DA neuron are about 20 mV above the corresponding values for the HH neuron (Figures 2A-B, 3A). Also, the DA neuron steady state functions are steeper than those of the HH neuron. The timescales of all three variables change by an order of magnitude as the system evolves in the phase space. Thus, there is no permanent timescale separation among the variables, and we only compare the timescales of the corresponding variables in the two neurons. The time constants of the $K^+$ and the $Na^+$ currents display steeper voltage dependence in the DA neuron and are an order of magnitude greater. This makes the $K^+$ current activation and the $Na^+$ current inactivation effectively slower than in the HH neuron (Figures 2C, 3B). The time constant $\tau_v$ of the membrane potential $v$ depends on what currents are open. Its minimum is determined by the conductance of the sodium current $\tau_{v\min} = C/g_{Na}$, and has a similar value in both neurons. Its maximum is approximated by the leak conductance $\tau_{v\max} = C/g_L$, and has a much greater value in the DA neuron ($\tau_v = 20$ ms) than in the HH neuron ($\tau_v = 3.3$ ms).

Below we change half-(in)activation parameter values $v_{nh}$ and $v_{hh}$ simultaneously with $\theta_n$ and $\theta_h$, respectively. These parameters are linked for all channels, and such manipulation is the most physiologically relevant.

**Results**

The DA neuron demonstrates Class 3 excitability [12]: The resting state remains stable for any value of the applied current. The oscillatory solution emerges from a saddle-node bifurcation of limit cycles, and stays completely isolated from the equilibrium state (Figure 4A). When the half-activation of the $K^+$ current, $v_{nh}$, is increased to -31 mV (Figure 4B), the class of excitability of the DA neuron changes to Class 2: The oscillatory solution emerges again from a saddle-node bifurcation of limit cycles, but in this case the equilibrium state becomes unstable via subcritical

Andronov-Hopf bifurcations. We use the parameter set from Figure 4B in all two-parameter bifurcation diagrams for the DA neuron given below.

The HH neuron possesses no hysteresis and has class 2 excitability (Figure 4C). However, relatively weak bistability and hysteresis compared to the DA neuron may be induced in the HH neuron with a decrease in the half-activation of the $K^+$ current $v_{nh}$ (Figure 4D). Similarly, we use the parameter set from Figure 4D in all two-parameter bifurcation diagrams for the HH neuron that follow.

**Half-(in)activation parameters' effect on hysteresis**

To investigate the effect of the half-(in)activation parameters on hysteresis in both neurons we consider the two-parameter bifurcation diagrams where the $K^+$ current half-activation $v_{nh}$ or the $Na^+$ current half-inactivation $v_{hh}$ are varied together with the applied current.

The two-parameter bifurcation diagram in $v_{nh}$ and $I_{app}$ for the DA neuron is shown in Figure 5A. This diagram shows the location of the Andronov-Hopf and the saddle-node bifurcations marked in Figure 4B for different values of the bifurcation parameters. Every horizontal cross section of this diagram at a particular value of $v_{nh}$ defines the instability range and the oscillatory range which are bounded by these bifurcations. The ranges extend in the parameter $v_{nh}$ and span 2-dimensional regions. Likewise, the bistability range spans the shaded region in the bifurcation diagram. We characterize the strength of hysteresis by the relative size of this shaded region compared to the instability region (bounded by solid curves in Figure 5).

At intermediate values of the applied current, the depolarization block boundary consists of two transitions: First, the equilibrium state becomes stable in a subcritical Andronov-Hopf bifurcation; second, the stable oscillatory solution disappears in a saddle-node bifurcation of limit cycles (see e.g. Figure 4B). In the range between the two bifurcations, the system is bistable and may show oscillations or a steady voltage depending on the initial conditions. At higher values of $v_{nh}$, the two bifurcation curves stay at a nearly constant distance (Figure 5A) and hysteresis remains strong. Instability range shortens and disappears at $v_{nh} = -15$ mV and oscillatory range follows at $v_{nh} = 14$ mV, at which point all oscillations cease in the DA neuron.

Figure 5D shows the same bifurcation diagram for the HH neuron. The area between the bifurcation curves is small. The model can show a significant bistability in response to variations in the applied current, but only with a very precise tuning of $v_{nh}$ to values right above -60 mV (see e.g. Figure 4D). The comparison of the areas of the bistability regions in the two neurons allows us to say that bistability is much stronger in the DA neuron.

An increase in the $Na^+$ current half-inactivation reduces and then completely abolishes hysteresis in both neurons (Figures 5B, E). In Figure 5B we fix the half-activation of the $K^+$ current at the level indicated in Figure 5A. Thus, Figure 5A and Figure 5B are perpendicular sections of the parameter space that intersect along the indicated levels. The influence of the $Na^+$ current half-inactivation is remarkably similar in the two neurons. Along with removing bistability, increasing half-inactivation expands the oscillatory range very much, so that it becomes similar in the two neurons (Figures 5B, 5E). This parameter change makes the inactivation effectively weaker, and the $Na^+$ window current greater. Therefore, a weaker $Na^+$ current inactivation promotes oscillations at a higher applied depolarization. However, a further elevation of the half-inactivation blocks oscillations completely. Thus, inactivation is necessary for generating oscillations in both neurons. Altogether, there is an optimal value of the half-inactivation of the $Na^+$ current that maximizes the oscillatory range and abolishes hysteresis in the model.

The influence of the $Na^+$ current half-activation $v_{mh}$ on the transition to the depolarization block in the DA and HH neurons is shown in Figures 5C and 5F, respectively. In the DA neuron a decrease in the $Na^+$ current half-activation initially expands both oscillatory and instability regions and moderately increases hysteresis, but below $v_{mh}$ =-31 mV, the dependence is reversed. Further decrease in $v_{mh}$ results in simultaneous shortening of oscillatory and instability regions, but their upper boundaries remain almost parallel and hysteresis remains strong. In the HH neuron (Figure 5F), the oscillatory range also peaks at an intermediate level of $v_{mh}$ around -45 mV. Both low and high values of the half-activation abolish oscillations. However, hysteresis exists only in a very narrow range of the $Na^+$ current half-activation and is substantially smaller than in the DA neuron (Figure 5C).

**Half-(in)activation slope parameters' effect on hysteresis**

The slopes of the activation and inactivation functions differ substantially in the DA and HH neurons (Figures 2A, B and 3A). Therefore, we also estimate the effect of the slope parameters on hysteresis in both neurons.

We change the slope of the Na$^+$ current inactivation function from steep ($|S_h| = 4$) as in the DA neuron to gradual ($|S_h| = 7$) as in the HH neuron (see Figure 2B) in both neurons. The decrease in the slope (larger $|S_h|$) increased the instability range. However, it reduced and then completely abolished bistability in both neurons (Figures 6A, D). This occurs because the instability range expands strongly and merges with the oscillatory range as $S_h$ increases in both neurons. Interestingly, in the HH neuron instability region was not present until the value of the slope parameter was around $|S_h| = 6.5$, but then instability region quickly expanded removing hysteresis similarly to Figure 5D. As for the variations in the half-inactivation above, the influence of $S_h$ in the two neurons is remarkably similar. First, it removes the difference in the length of the instability region between the neurons. Second, its increase abolishes hysteresis. Therefore, the slope of the Na$^+$ current inactivation controls hysteresis and the length of the oscillatory and instability ranges in both neurons.

The Na$^+$ current activation function is steeper in the DA than in the HH neuron (Figure 2A). A decrease in the slope of the function (increasing $S_m$ from 8 to 9, Figure 6B) leads to an almost linear expansion of the instability and oscillatory regions in the DA neuron. Hysteresis remains almost unchanged because the boundaries are parallel. In the HH neuron, both the instability and oscillatory regions shorten and shift into higher values of the applied current with a steeper activation of the Na$^+$ current ($S_m \to 8$). While hysteresis increases at the low applied currents (hyperpolarization block), at the depolarization block, hysteresis remains unchanged because upper boundaries of both the oscillatory and instability ranges expand almost equally. Altogether, changes in $S_m$ only slightly shift the upper boundaries, and both the oscillatory and the instability ranges remain of different orders of magnitude in the HH and DA neurons. Moreover, the boundaries shift in the opposite directions in the two neurons, which only emphasizes the difference between their parameter sets.

To test how steepness of the activation function for the K$^+$ current affects bistability we change the slope of the K$^+$ current activation in both neurons from steep ($S_n$ = 8) as in the DA neuron to a more gradual ($S_n$ = 15) as in the HH neuron (Figure 3A). In the DA neuron (Figure 6C), the upper boundaries of oscillatory and instability regions remain almost parallel until the instability region disappears at around $S_n$ = 12. Even above that point, the slope parameter only weakly affects the boundary of the oscillatory region, and hysteresis remains almost unchanged. In contrast, in the HH neuron (Figure 6F) the oscillations are not present for smaller values of $S_n$, i.e. for the steeper voltage dependence of the K$^+$ current. The oscillatory range emerges above $S_n$ = 10 and quickly expands, whereas the equilibrium remains stable giving rise to strong hysteresis. The instability region appears above $S_n$ = 13, rapidly expands and limits hysteresis to short ranges at both boundaries. Therefore, the decrease in the slope of the K$^+$ current activation function ($S_n \rightarrow$ 15) has an opposite effect on the instability range in the two neurons. Extending the distinction, in the HH, but not in DA neuron, the parameter strongly affects the oscillatory and hysteresis regions.

**Gating variables' kinetics effect on hysteresis in the DA and the HH neurons**

None of the variables in the model is uniformly slow or uniformly fast. The timescale of the voltage is minimal when the Na$^+$ current is open, and elevates to the maximum when the leak current works alone. Likewise, the timescales of the gating variables depend on the voltage (Figures 2C and 3B). Therefore, there is no permanent separation onto fast and slow variables in the model. On the other hand, comparing timescales of the corresponding variables in the two neurons is more straightforward and has a clear physiological meaning. By changing kinetics of the gating variables below, we study how the difference in the timescales affects the oscillatory and instability ranges.

The inactivation variable is much faster in the HH neuron compared to the DA neuron (Figure 2C). Therefore, we now study how hysteresis is affected by accelerating the Na$^+$ current inactivation uniformly at all voltages ($f_h$ > 1). Figure 7A shows that accelerating the inactivation moderately reduces the distance between the Andronov-Hopf and the saddle-node of limit cycles bifurcations. Instability range shortens slower than the oscillatory range. After the instability range disappears, the oscillatory range continues to shorten until all oscillations cease in the DA neuron. In the HH neuron, the transition should be made in the opposite direction because inactivation is initially much faster compared to the DA neuron; therefore we reduce its rate of change ($f_h < 1$). In contrast to the DA neuron, the instability range does not depend on the inactivation timescale (Figure 7D). The oscillatory range shortens with slower $Na^+$ current inactivation because the hysteresis ranges at both boundaries of the instability range disappear (Figure 7D).

Similar results hold when the $K^+$ current activation variable $n$ is accelerated ($f_n > 1$) in both the DA neuron and the HH neuron (Figures 7B, 7E). In the DA neuron, instability range shortens and disappears at around four times faster activation ($f_n = 4$) of the $K^+$ current. The oscillatory range, first, expands with a faster $K^+$ activation, but then starts to decrease in size. This decrease is more gradual than in Figure 7A for the $Na^+$ current inactivation, and bistability remains in the neuron when the $Na^+$ current becomes as fast as in the HH neuron ($f_n = 10$). In the HH neuron, the instability range weakly depends on the timescale of the $K^+$ current activation variable $n$. When the variable becomes slower ($f_n < 1$), the oscillatory range shortens, and hysteresis disappears ($f_n = 0.1$).

Finally, we change both the $Na^+$ current inactivation and the $K^+$ current activation timescales simultaneously in both neurons. The diagram for the DA neuron (Figure 7C) is very similar to Figure 7A. This suggests that the changes in the dynamics are not due to the introduced mismatch between the timescales of the two gating variables, but mostly due to the mismatch between the timescales of the voltage and the gating variables. Furthermore, as follows from the similarity of Figures 7A and 7C, the accelerated $Na^+$ current inactivation contributes the most to the loss of oscillations. When $K^+$ current activation is also accelerated, the gating variables remain at the same timescale, but this only moderately expands the oscillatory region.

In the HH neuron, when we make the kinetics of both variables slower ($f_{n,h} \to 0.1$), hysteresis at the upper boundary of the oscillatory range increases (Figure 7F). This is opposite to the results for the differential changes in these two parameters above (Figures 7D, E). This also contrasts the results for the DA neuron. Hence, the reduction in hysteresis was due to a mismatch between the timescales of the gating variables. By contrast, concurrent slowing of the gating variables, which creates a mismatch between the timescale of voltage and that of the two gating variables, expands the hysteresis range.

We think that the changes in the hysteresis range are due to the effective dimensionality of the model. Suppose that the model is effectively 2-dimensional, that is, one of the three variables passively follows the other two. By introducing a timescale separation, we make the oscillator of relaxation type. A two-dimensional relaxation oscillator (e.g. FitzHugh-Nagumo) does not display significant hysteresis. There are geometric reasons for that: the equilibrium state is very close to the limit cycle at the bifurcation transition. However, if the relaxation oscillator is (effectively) 3-dimensional and have one fast and two slow variables, hysteresis can be made much more pronounced. The same geometric reasoning can lead to this conclusion because the limit cycle and the equilibrium state can now be separated more in 3 dimensions. Our results suggest that the two gating variables are required to be equally slow compared to the voltage in order to enhance hysteresis in the HH neuron. This is the combination that produces a relaxation oscillator with one fast and two slow variables.

**Contribution of other parameters to hysteresis**

The susceptibility of the DA neuron to depolarization block was attributed to the weakness of the delayed rectifier current long ago. The common sense explanation is that the voltage stays high near the state of depolarization block because the potassium current cannot lower it enough. The increase in the maximal conductance of the $K^+$ current $g_K$ in the DA neuron leads to the monotone increase of the oscillatory range and the decrease and disappearance of the instability range (Figure 8A). The growth of the oscillatory range is consistent with the logic outlined above, but stabilization of the equilibrium state that entail strong hysteresis is unexpected. In the HH neuron, the increase in $g_K$ also leads, at first, to the expansion of the instability range without hysteresis. Then the instability range shortens abruptly, but the oscillatory range persists, and a significant hysteresis region emerges (Figure 8D). Hence, the maximal conductance of the $K^+$ current $g_K$ efficiently controls the length of the oscillatory range in both neurons. However, the strength of hysteresis is controlled by other parameters because it is drastically different in the two diagrams.

The reversal potential of the K$^+$ current is defined by the extracellular potassium concentration, can be controlled in experiments, and has been found to affect hysteresis in the model of pre-Botzinger complex respiratory neuron (Y. Molkov, private communications). An increase in E$_K$ monotonically reduces oscillatory and instability ranges in both the DA and the HH neurons (Figures 8B, E). Furthermore, the upper boundaries of both oscillatory and instability regions in the HH neuron are very close to straight parallel lines (Figure 8E). This suggests a passive contribution of the K$^+$ current at the transition to depolarization block and a linear compensation by $I_{app}$. In the DA neuron, this dependence is more nonlinear (Figure 8B), and the reduction in hysteresis is much stronger.

The maximum conductance of the leak current is very different in the two neurons (see Table 1), and determines the slowest timescale of voltage changes (see discussion in the Model subsection). The increase in $g_L$ in the DA neuron shortens both instability and oscillatory ranges (Figure 8C). Instability range disappears first and then the oscillatory range follows at $g_L = 0.4$. This influence of increasing $g_L$ is very similar to the influence of accelerating both gating variables (compare Figures 8C and 7C). This is counterintuitive because increasing $g_L$ is equivalent to accelerating the membrane potential $v$ and, therefore, should have had a similar effect to slowing $n$ and $h$. This differentiates the oscillatory mechanism in the DA neuron from a relaxation oscillator. In a relaxation oscillator, oscillations disappear if the timescale separation is decreased or reversed. Therefore, accelerating a slow variable would abolish oscillations, whereas accelerating a fast variable would only promote them. The fact that accelerating any variable abolished oscillations in the DA neuron means that the oscillatory mechanism does not tolerate a significant mismatch in the timescales (hence, similarity between Figures 8C and 7C), distinguishing it from the relaxation oscillator.

In the HH neuron, increase in the maximum conductance of the leak current shortens both the oscillatory and instability ranges (Figure 8F). This also tells against the relaxation oscillator mechanism and the role of the voltage as a fast variable in the HH neuron. The elevation in $g_L$ significantly increases the hysteresis range. Increasing $g_L$ makes the voltage, or more precisely its slowest timescale, faster. This introduces timescale separation similar to the one achieved by slowing the two gating variables (Figure 7F). In both cases (Figures 7F and 8F) this leads to the increase in hysteresis at the depolarization block. However, slowing the gating variables hardly affects the instability region and only reduces hysteresis at the hyperpolarization block. Therefore, the increase in the maximum of the voltage timescale determined by the leak conductance promotes oscillations and the instability of the equilibrium state in both neurons.

**Normalized contributions of parameters to hysteresis**

To compare the effect of different parameters on hysteresis we compute changes in the hysteresis range with changes in each parameter. The parameters were increased by 10%. The changes in the length of the hysteresis range were normalized by its initial length to obtain the relative contributions. The results are shown in Table 2. For example, the increase in the maximal conductance of the $K^+$ current in the HH neuron from 36 mS/cm$^2$ to 39.6 mS/cm$^2$ (10% increase) leads to the increase in the length of hysteresis range from 30.5 to 67.8 (122% increase) (see Figure 8D). The increase in the (in)activation parameters $v_{hh}$ and $v_{nh}$ in the HH neuron leads to the complete removal of hysteresis (Figures 5C, D) and is reflected by the 100% decrease in hysteresis in Table 2. Overall, normalized change in hysteresis in the HH neuron is an order of magnitude higher than in the DA neuron for most of the parameters. This provides another indication that hysteresis in the HH neuron exists in the narrow parameter ranges and is not as robust as in the DA neuron since small changes in parameter values lead to large changes in hysteresis.

**Summary of the results**

The parameters in the two model neurons can be separated into three groups. The first group of parameters consists of the half-inactivation of the $Na^+$ current $v_{hh}$ and the slope of the inactivation function $S_h$. The two-parameter diagrams in $v_{hh}$ and $S_h$ are very similar for the HH and the DA neurons in spite of the differences in other parameters (Figures 5B, E and 6A, D). Not only these parameters control hysteresis in a similar way, but the two-parameter diagrams show the strongest expansion of the oscillatory and instability ranges. By changing the two parameters, we can remove the order of magnitude difference in the length of these ranges in the HH vs. DA neuron. Therefore, these parameters contribute most to the difference between the neurons in both hysteresis and the length of the instability/oscillatory ranges.

The second group of parameters includes the half-activation of the $K^+$ and $Na^+$ currents $v_{nh}$ and $v_{mh}$, reversal potential of the $K^+$ current $E_K$, leak conductance $g_L$ and the slope of the activation function of the $Na^+$ current $S_m$. Variations in these parameters produce the diagrams that are quite distinct in the HH and DA neurons, i.e. cannot make the dynamics of the two neurons similar. This suggests that these parameters do not contribute to the difference between the DA and the HH neurons. Nevertheless, they influence hysteresis and the length of the oscillatory and instability ranges similarly in both neurons.

Finally, the third group of parameters consists of the slope of the $K^+$ current activation function $S_n$ and parameters that influence the kinetics of the gating variables, i.e. $f_n$, $f_h$ and $f_{n,h}$. Variations in these parameters not only produce different diagrams, but also influence the neurons in the opposite ways.

*Discussion*

In this paper, we have analyzed bistability that distinguishes two types of neurons. We identified their spike-producing currents as responsible for bistability. The models were reduced to the same system and differed by the values of the parameters. We examined transitions between the two parameter sets and found that bistability is present in a wide region of the multidimensional parameter space. The values of the parameters in the bistability regions are physiologically plausible because transitions span the intervals between values corresponding to two types of neurons. This is consistent with bistability between tonic spiking and the silent state commonly observed in neurons. Our modeling suggests that this bistability arises from the interaction of the spiking currents.

Bistability is useful in qualitative classification of neurons based on the firing patterns. Electrophysiological and pharmacological characterization of neurons separates them into numerous types. The neurons differ by their neurotransmitters, the composition of currents, typical firing patterns, responses to pharmacological manipulations, etc. Managing the diversity of neurons is an enormously complex task. Thus, of critical importance are criteria that can identify broad classes of neurons sharing some functional similarities. A great example is the neuron characterization by a phase response curve [16] or classification of neurons into resonators and integrators; bistable and monostable dynamical systems [12]. These characteristics separate broad groups of neurons, and are very useful for predicting how neurons behave when they interact in a network. Another example is the separation of neurons into three classes of excitability [17]. The class of neuron excitability is determined in one of the simplest experiments – a negative current is applied into the soma through an electrode and then gradually removed. In response, the neuron first enters the silent state of hyperpolarization and then resumes firing as the hyperpolarizing applied current is removed. The transition from quiescence to firing determines the excitability class.

Bistability between tonic spiking and silence has been used in explaining the mechanisms of bursting [12]. In this case, an additional variable plays the role of a parameter that provides hysteresis and switches the system from spiking to silence and back. Only artificially treating this variable as a parameter in the model allows for observing bistability in simulations. We consider a true parameter, applied current, and bistability that occurs in experiments as the parameter is manipulated. Our model did not take into account the subthreshold currents. Their inclusion may suppress bistability in some cases. Our modeling of the DA neuron [14] shows that a model that includes subthreshold currents together with the spike-producing ones retains the same bistability. How bistability is affected by subthreshold currents in other neurons is a subject of future studies focused on particular neurons.

We have found that bistability is much stronger in the DA neuron than in the HH neuron. The major factors contributing to this difference are a low half-inactivation and a steep voltage dependence of the inactivation of the $Na^+$ current. Only the manipulations of these two parameters were able to abolish the order of magnitude difference in the length of the oscillatory region in the two neurons. They also control hysteresis in a very similar way in spite of the difference in other parameters. The rest of the parameters produce very different diagrams in the two neurons. Some of them have the opposite influence on the dynamics of the two neurons.

In order to interpret the result for the future experiments, we connect it to physiological characteristics. The window Na$^+$ current is its small steady state component that remains after a strong transient component as the current inactivates. Lowering half-inactivation parameter of the current decreases its window component. Increasing the slope of the inactivation voltage dependence reduces the window current as well. Our way of changing the slope excludes any shift in the half-inactivation. Thus, two manipulations that decrease the window current promote bistability. Altogether, our results suggest a connection between the characterization of the Na$^+$ window current in a neuron and strong bistability in response to changes in the applied current.

Bistability endows neurons with richer forms of information processing. A bistable cell encodes a brief signal by a long-lasting change in its firing. Hence, the bistability between resting and tonic spiking states studied in this article has been hypothesized to be involved in short-term memory (discussed in [7]). This type of bistability was also observed in different motor neurons [4-6]. Bistable motor neurons have been hypothesized to support prolonged low force tasks, like posture.

Hysteresis at the upper boundary of oscillatory range may be essential for pacemaker-type neurons as it may improve robustness of oscillations and lead to a more efficient control of the dynamics [9]. Efficiency and robustness follow from the inability of small perturbations (e.g. noise) in the control parameter to switch the neuronal activity from one mode to the other as soon as the bifurcation parameter (applied current) is perturbed. For this reason, many physical systems like heating thermostats utilize hysteresis to improve efficiency by reducing the frequency of on-off switching.

Bistability studied in this article is generally independent of the excitability class. In particular, Class 3 excitability is always accompanied by bistability simply because the equilibrium state remains stable for the whole parameter range where the oscillatory solutions exist. However, in other cases where the equilibrium loses stability, the excitability class is unrelated to the presence of bistability. In most cases we examined, strong bistability occurs at the upper boundary of the instability (oscillatory) range, i.e., at the depolarization block. The excitability class refers to the transition at the hyperpolarization block. These two transitions are independent. The saddle-node on invariant circle bifurcation, which is responsible for Class 1 excitability, never occurs at the upper boundary of the oscillatory range. The half-activation of the $K^+$ current was the most effective in spanning all three excitability classes, but did not abolish bistability completely. By contrast, changing the half-inactivation of the $Na^+$ current or kinetics of the gating variables switches the excitability only between Class 2 and Class 3. Therefore, bistability can be used in conjunction with the excitability class in characterizing the neurons.

All together, our paper prepares a theoretical basis for experimental studies of bistability: bistability at the depolarization block should be very common among neurons, and we have determined what it characterizes, as discussed above. The implications of this hysteresis may be as significant as the enhanced therapeutic effect of antipsychotic drugs (Figure 1). Should the terminology be corrected in order to prepare for the future experiments? Historically, suppression of oscillations with a growing applied current is called depolarization block. This corresponds to the upper boundary of the oscillatory range and only partially characterizes the transition. This is the boundary observed in experiments when the applied current increases, but not when it decreases over the same range. The other part is the upper boundary of the instability range. We can introduce a name for this transition; however, bistability is much more important than the transition itself, and we suggest to directly measure the bistability range in experiments.

**Figure Captions**

**Figure 1. Simulated effect of antipsychotic drugs in the DA neuron.** The influence of the drugs is modeled as excitation by applied current (lower trace) Tonic firing in the DA neuron (upper trace) is interrupted with excessive excitation. DA neuron remains silent after complete withdrawal of the excitation due to hysteresis. Parameters for the DA neuron are from Table 1.

**Figure 2. The activation A) and inactivation B) functions of the Na$^+$ current in the DA neuron (solid curve) and the HH neuron (dashed curve).** C) The time constant function of the Na$^+$ current. Note that the functions are shifted by around 20 mV for a better comparison of the slopes. The ranges for the HH neuron are at the top and to the right.

**Figure 3. The activation A) and time constant B) functions of the K$^+$ current from the DA neuron (solid curve) and the HH neuron (dashed curve).** The ranges for the HH neuron are at the top and to the right.

**Figure 4. One-parameter bifurcation diagrams for the DA neuron and the HH neuron.** Hysteresis at the upper boundary of the oscillatory range (where it exists) is indicated by arrows showing direct and reverse transitions. A) Oscillatory solution stays isolated from the equilibrium state in the DA neuron. This is the Class 3 excitability. Parameters for the DA neuron are from Table 1. B) The oscillatory solution connects to the equilibrium state in an Andronov-Hopf bifurcation. Parameters are from A), except that the K$^+$ current half-activation is increased by 4 mV ($v_{nh}$ = -31 mV). C) Hysteresis is not present at the upper boundary of the oscillatory range in the HH neuron. Parameters for the HH neuron are from Table 1. D) The oscillatory solution connects to the equilibrium state in an Andronov-Hopf bifurcation. Parameters are from C), except that the K$^+$ current half-activation is decreased by 5 mV ($v_{nh}$ = -58 mV). Thin curves represent equilibrium states, thick curves - limit cycles. Solid (dashed) curves represent stable (unstable) solutions. HB is the Andronov-Hopf bifurcation, SNLC is the saddle-node of limit cycles bifurcation.

**Figure 5. Two-parameter bifurcation diagrams of the DA neuron and the HH neuron in $v_{nh}/v_{hh}/v_{mh}$ and $I_{app}$ planes.** The hysteresis regions are shaded gray. A) Hysteresis is strong in the DA neuron. Parameters are from Table 1. B) Hysteresis is removed and the oscillatory region

expands in the DA neuron with the increase in the half-inactivation of the $Na^+$ current. Parameters are from Figure 4B. C) Hysteresis is not reduced with the decrease in the half-activation of the $Na^+$ current. Parameters are from Figure 4B. D) Hysteresis is weak in the HH neuron. Parameters are from Table 1. E) Hysteresis is removed in the HH neuron with an increase of $v_{hh}$. Parameters are from Figure 4D. F) Hysteresis is weak in the HH neuron with an increase of $v_{mh}$. Parameters are from Figure 4D. A solid curve represents an Andronov-Hopf bifurcation, a dashed curve – a saddle-node bifurcation of limit cycles. Horizontal dotted lines in A) and D) represent the values of half-activation of the $K^+$ current taken in B), C) and E), F), correspondingly. Horizontal dotted lines in B), C) and E), F) represent the values of half-(in)activations from Table 1 for the DA and HH neurons, correspondingly.

**Figure 6. Two-parameter bifurcation diagrams of the DA and HH neuron for the change in slope factors of (in)activation functions.** The hysteresis regions are shaded gray. A) A gradual voltage dependence of the $Na^+$ current inactivation function removes hysteresis in the DA neuron. B) Steeper voltage dependence of the activation of the $Na^+$ current has almost no effect on hysteresis in the DA neuron. C) More gradual voltage dependence of the $K^+$ current has little effect on hysteresis in the DA neuron. Parameters are from Figure 4B. D) More gradual voltage dependence of the $Na^+$ current reduces and then completely abolishes hysteresis in the HH neuron. E) Gradual voltage dependence of the activation of the $Na^+$ current has little effect on hysteresis at the upper boundary of oscillatory region in the HH neuron. F) Hysteresis range peaks at intermediate values of $S_n$ in the HH neuron. A solid curve represents an Andronov-Hopf bifurcation, a dashed curve – a saddle-node bifurcation of limit cycles. Horizontal dotted lines in A) and D) mark the value of slope parameter from Table 1 for the HH neuron.

**Figure 7. Changing kinetics of gating variables in the DA neuron and the HH neuron.** The hysteresis regions are shaded gray. A), B) Bistability range shortens with accelerating the gating variables kinetics in the DA neuron. $f_h=1$ and $f_n=1$ correspond to parameter set from Figure 4B. C) Simultaneous acceleration of both n and h variables decreases the size of bistability range. $f_{n,h}=1$ corresponds to parameter set from Figure 4B. D), E) Hysteresis is reduced or eliminated in the HH neuron with slowing the individual current kinetics. $f_h=1$ and $f_n=1$ correspond to parameter set from Figure 4D. F) Hysteresis is increased with simultaneous slowing of gating variables $n$ and $h$. $f_{n,h}=1$ corresponds to parameter set from Figure 4D. A solid curve represents an Andronov-Hopf bifurcation, a dashed curve – a saddle-node bifurcation of limit cycles. Horizontal dotted lines (where shown) give the values of $f_h$ and $f_n$ for which the maximum value of the corresponding time constant function for the DA (HH) neuron matches the maximum value of the time constant for the HH (DA) neuron.

**Figure 8. Two-parameter diagrams for the change in maximal conductances and equilibrium potential.** The hysteresis regions are shaded gray. A) Increase in $g_K$ expands the oscillatory region, shortens the instability region and increases hysteresis in the DA neuron. B) Increase in K$^+$ current reversal potential shortens oscillatory region much faster than the instability region, reducing hysteresis in the DA neuron. C) Increase in the maximum conductance of the leak current shortens both instability and oscillatory regions and finally eliminates oscillations in the DA neuron. D) Hysteresis exists in a narrow range of parameter $g_K$ in the HH neuron. E) Hysteresis at the upper boundary of the oscillatory region is not affected by the decrease in $E_K$ in the HH neuron. F) Hysteresis is slightly reduced with the decrease in the maximum leak conductance in the HH neuron. A-C) Parameter values from Figure 4B. D-F) Parameter values from Figure 4D. A solid curve represents an Andronov-Hopf bifurcation, a dashed curve – a saddle-node bifurcation of limit cycles. Horizontal dotted lines mark parameter values from Table 1 for the DA and the HH neurons, correspondingly.

**Table Captions**

**Table 1. Parameter values for the DA neuron and the HH neuron.**

| Parameter | Model | | Dimension |
|---|---|---|---|
| | DA neuron | HH neuron | |
| C | 1 | 1 | µF/cm$^2$ |
| $g_K$ | 4 | 36 | mS/cm$^2$ |
| $g_{Na}$ | 150 | 120 | mS/cm$^2$ |
| $g_L$ | 0.05 | 0.3 | mS/cm$^2$ |
| $E_K$ | -90 | -77 | mV |
| $E_{Na}$ | 55 | 55 | mV |
| $E_L$ | -34.4 | -54.4 | mV |
| Na$^+$ current activation constants | | | |
| $v_{mh}$ | -18 | -40 | mV |
| $S_m$ | 8 | 9 | |
| Na$^+$ current inactivation constants | | | |
| $v_{hh}$ | -48 | -62 | mV |
| $S_h$ | -4 | -7 | |
| $\tau_h^0$ | 1 | 1.2 | |
| $\tau_h^1$ | 55 | 7.4 | |
| $\theta_h$ | -53 | -67 | |
| $S_h^\tau$ | 12 | 20 | |
| DR current activation constants | | | |
| $v_{nh}$ | -35 | -53 | mV |
| $S_n$ | 8 | 15 | |
| $\tau_n^0$ | 5 | 1.1 | |
| $\tau_n^1$ | 51 | 4.7 | |
| $\theta_n$ | -79 | -53 | |
| $S_n^\tau$ | 23 | 50 | |

**Table 2. Normalized parameter contribution to hysteresis in the DA and HH neurons.**

| Model | Parameters | | | | | | | |
|---|---|---|---|---|---|---|---|---|
| | $g_K$ | $g_{Na}$ | $g_L$ | $v_{hh}$ | $v_{nh}$ | $S_h$ | $S_m$ | $S_n$ |
| HH neuron, % | 122 | -56 | 3 | -100 | -100 | 426 | -55 | -21 |
| DA neuron, % | 4 | 6 | 0 | 5 | 0 | 1.5 | -0.5 | 0 |

Initial parameter values for the DA and HH neurons were taken from Figures 4B and 4D, correspondingly.

Figure
Click here to download Figure: Fig1.eps

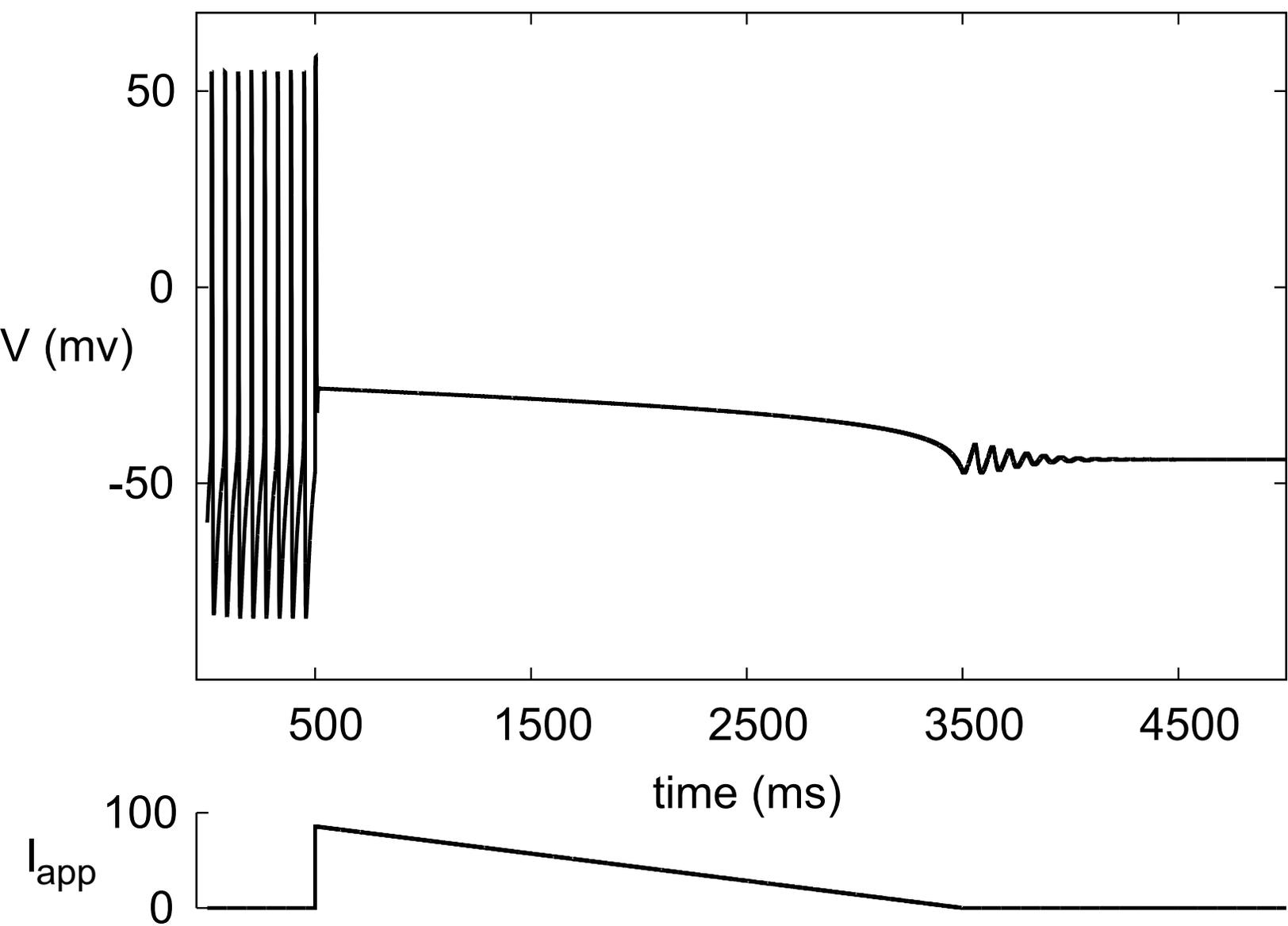

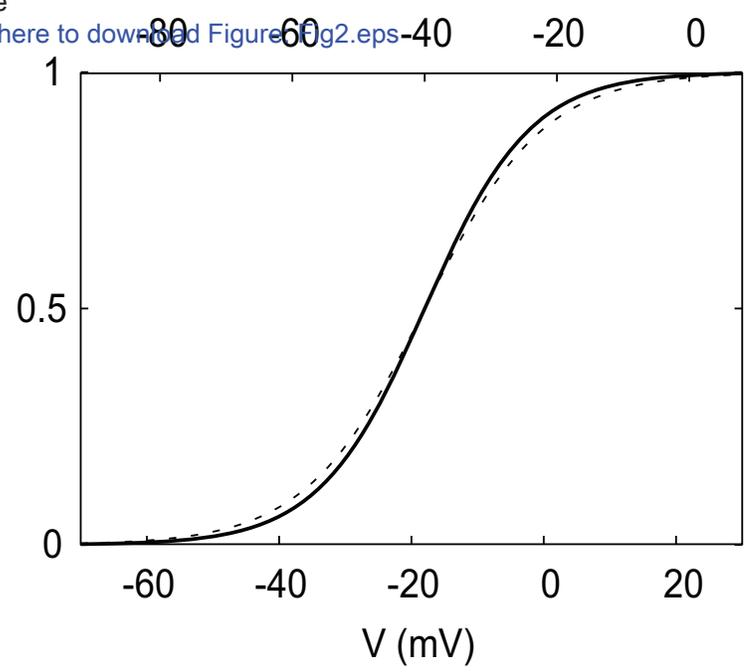
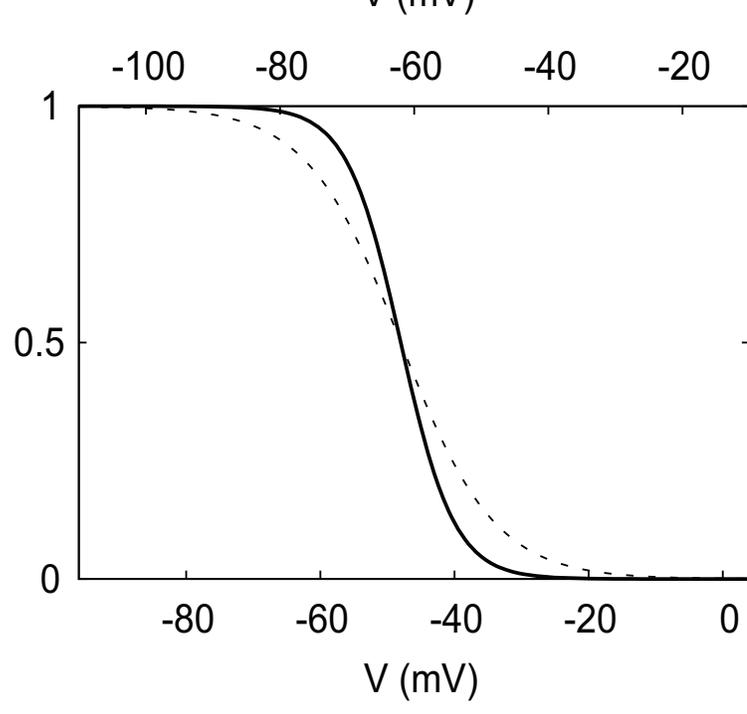
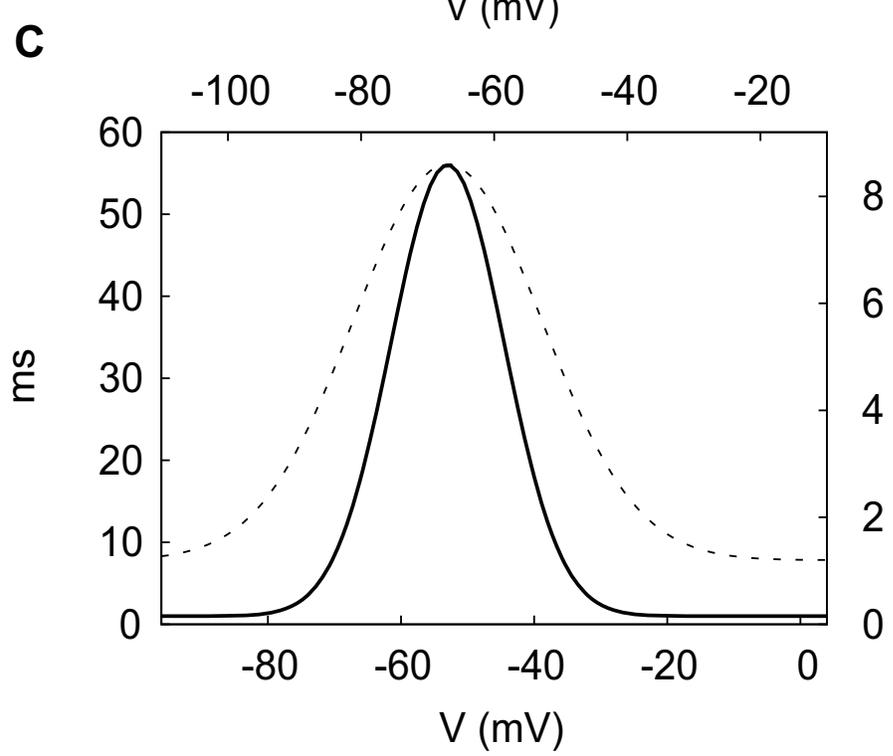

Figure
Click here to download Figure: Fig3.eps

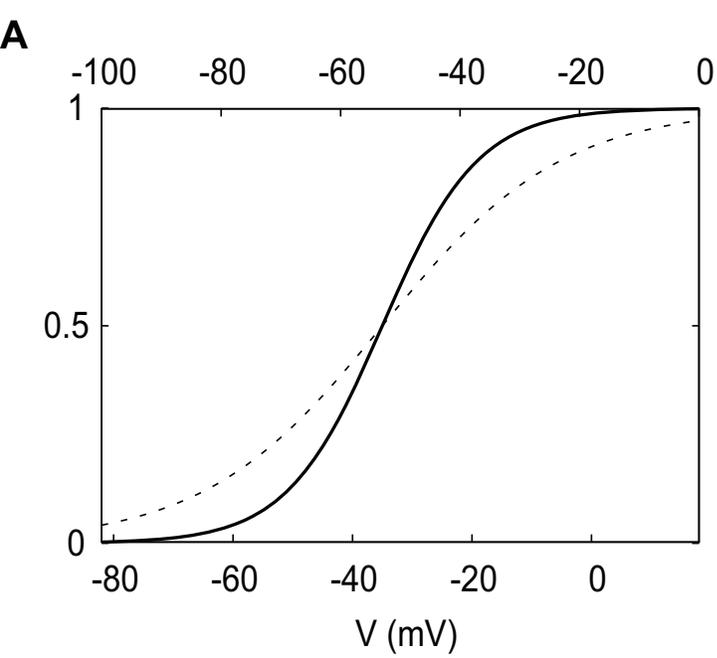

**A**

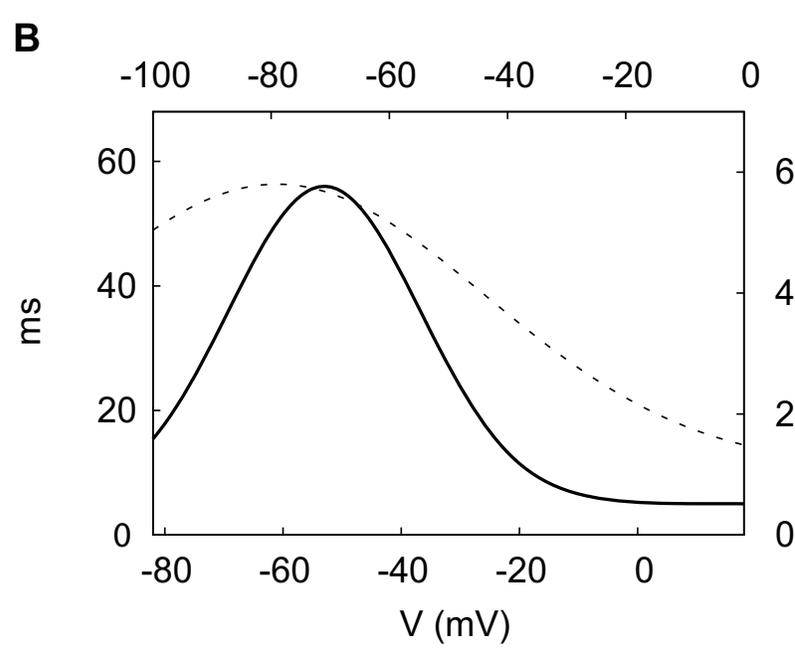

**B**

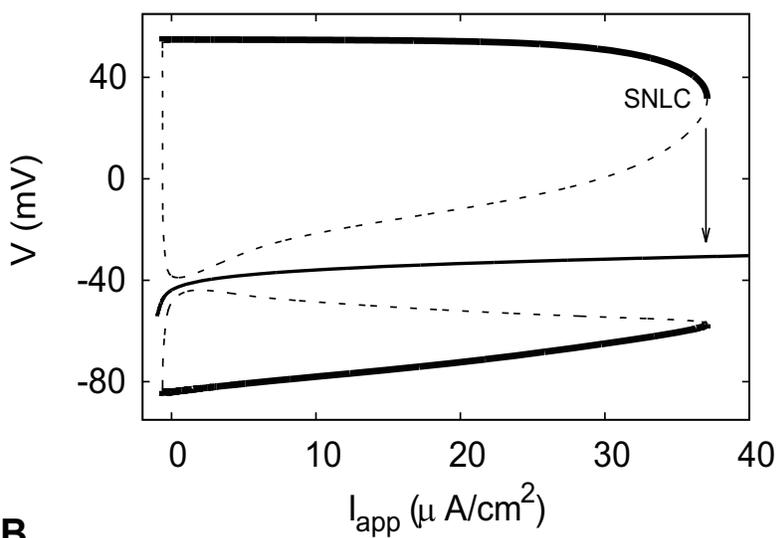
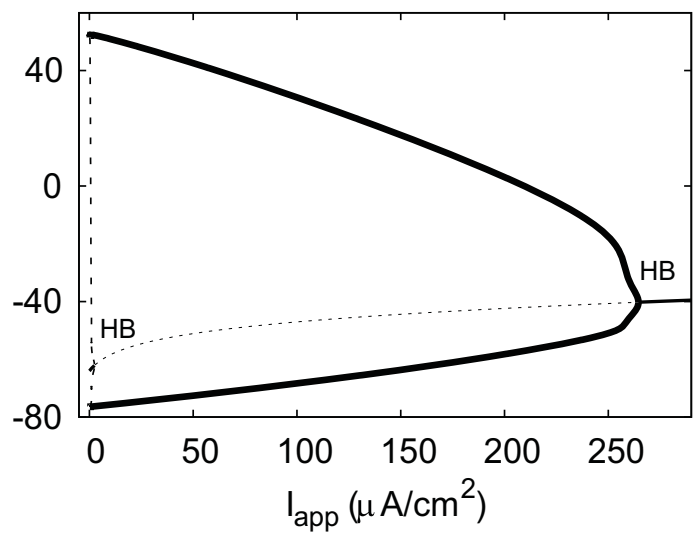
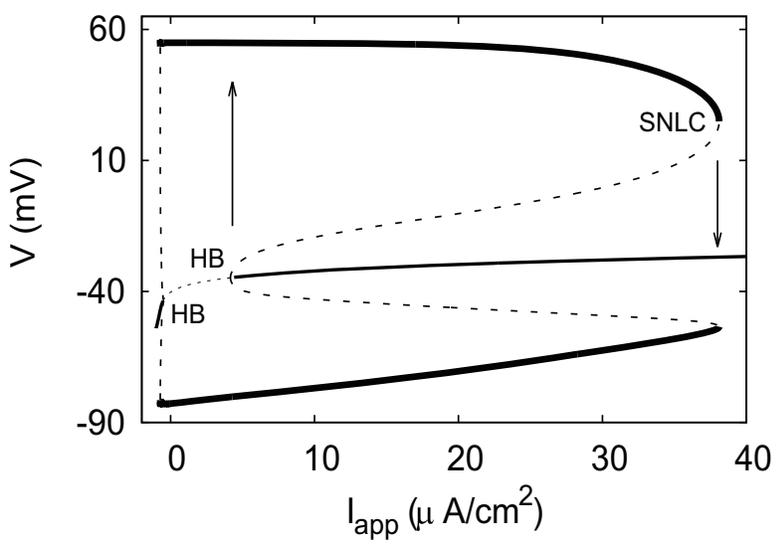
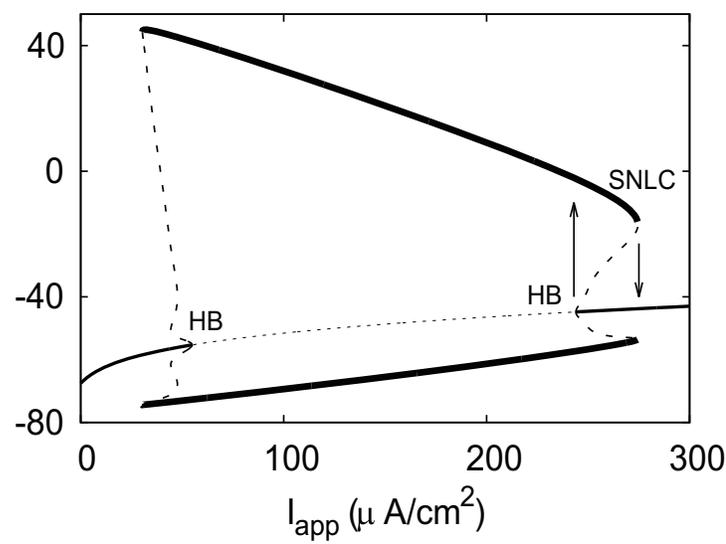

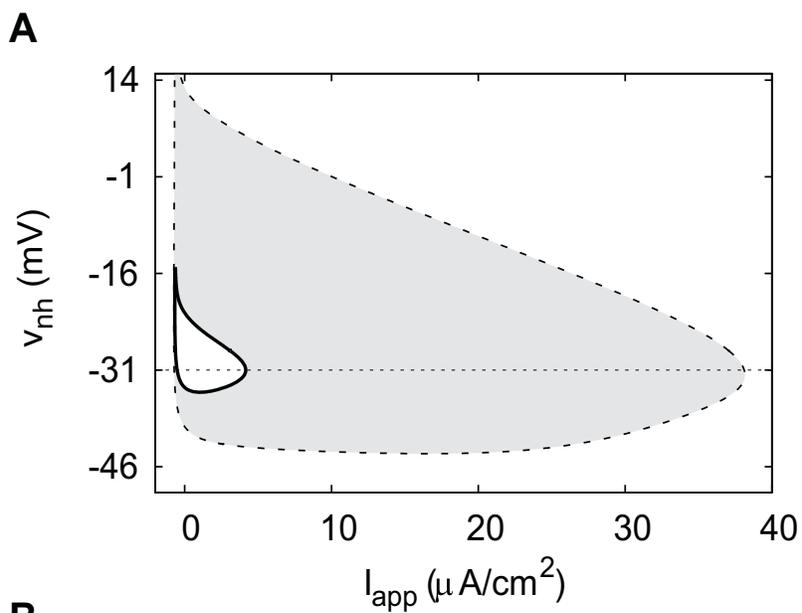
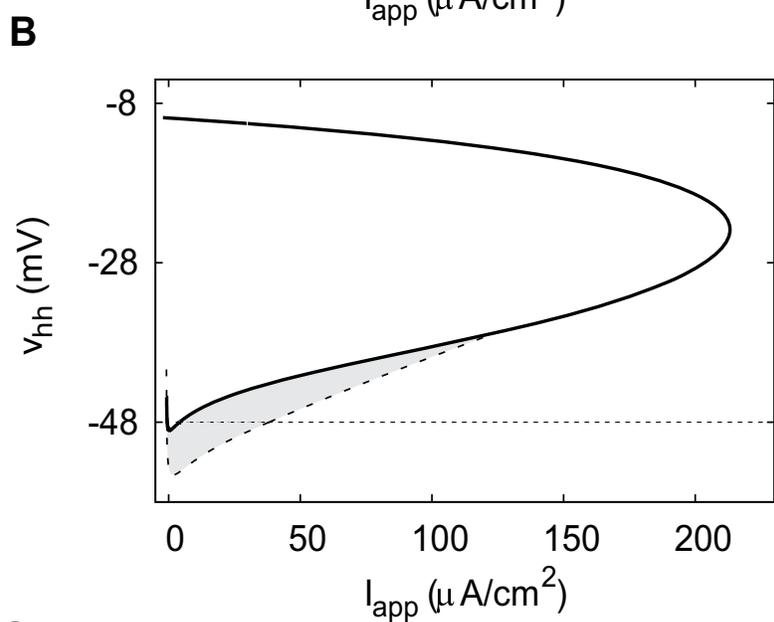
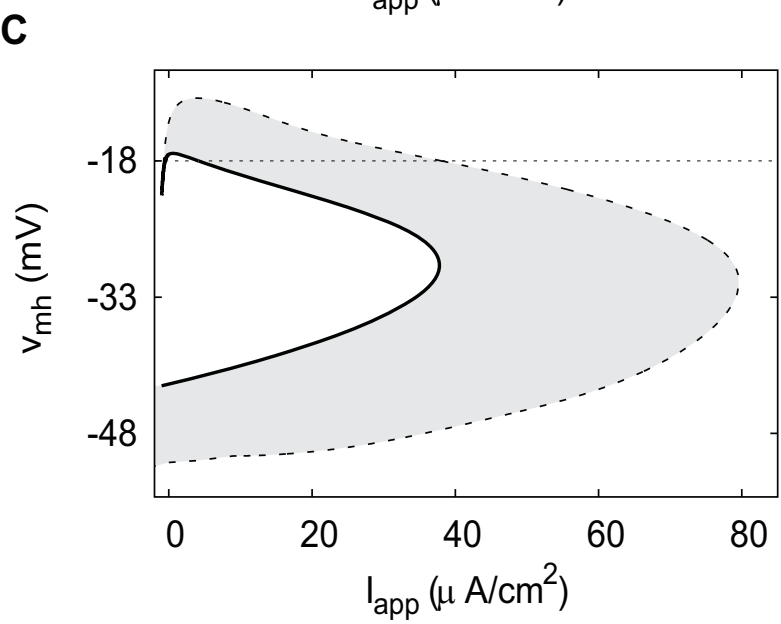
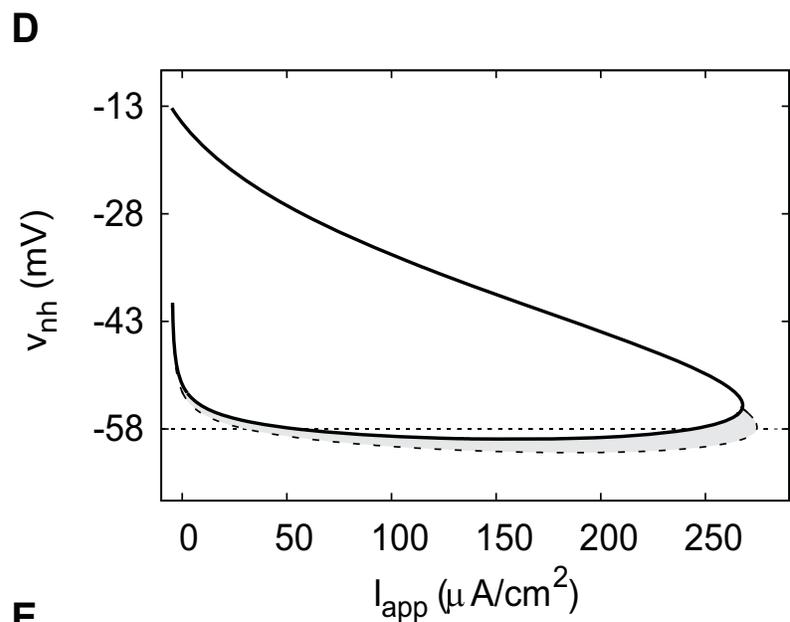
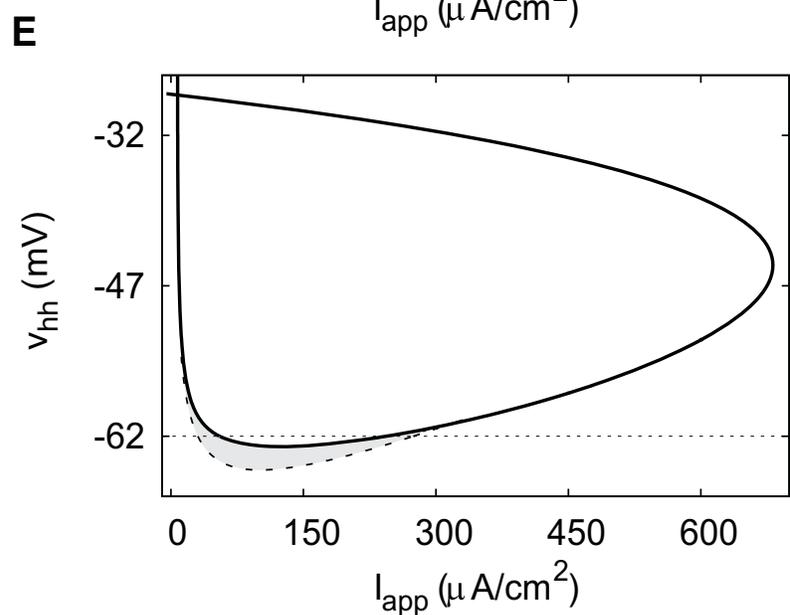
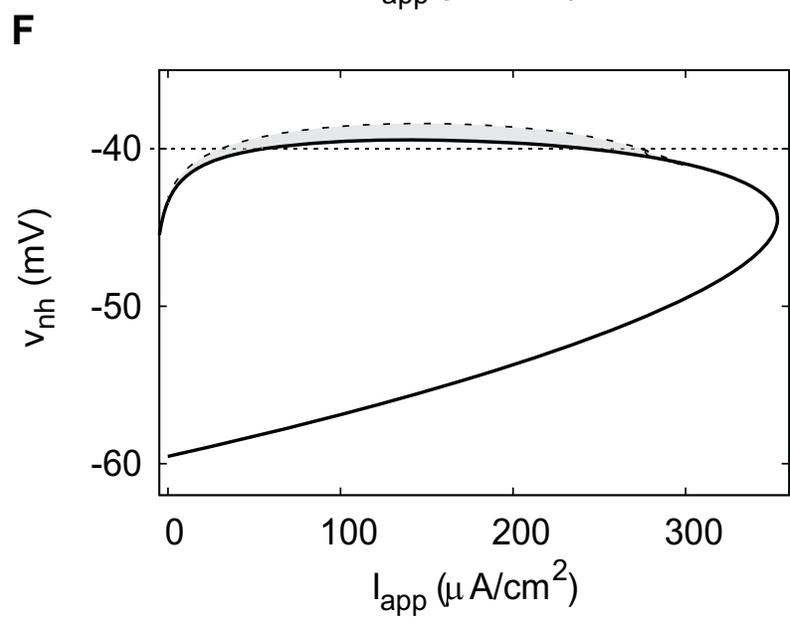

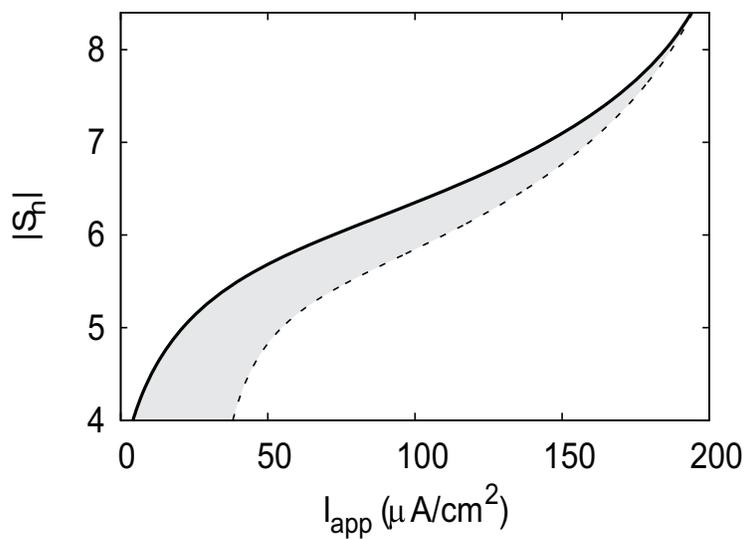
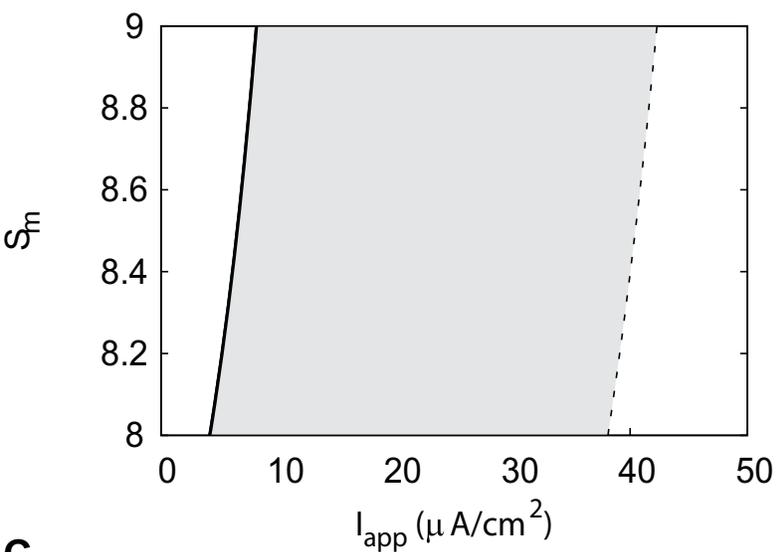
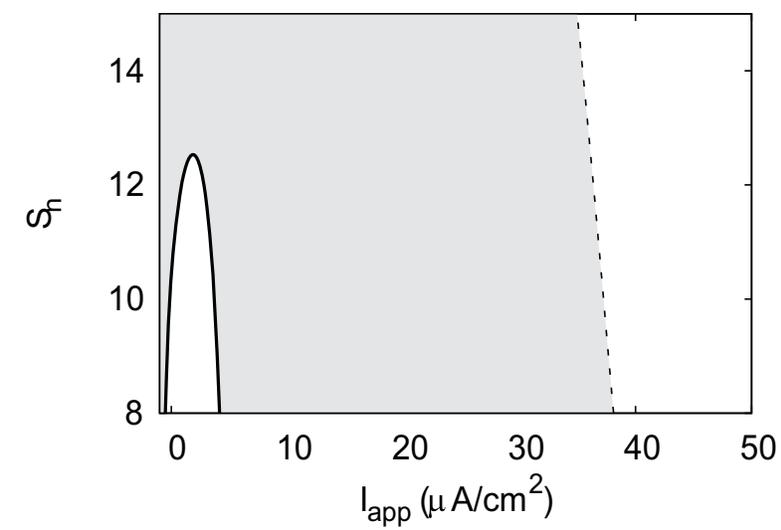
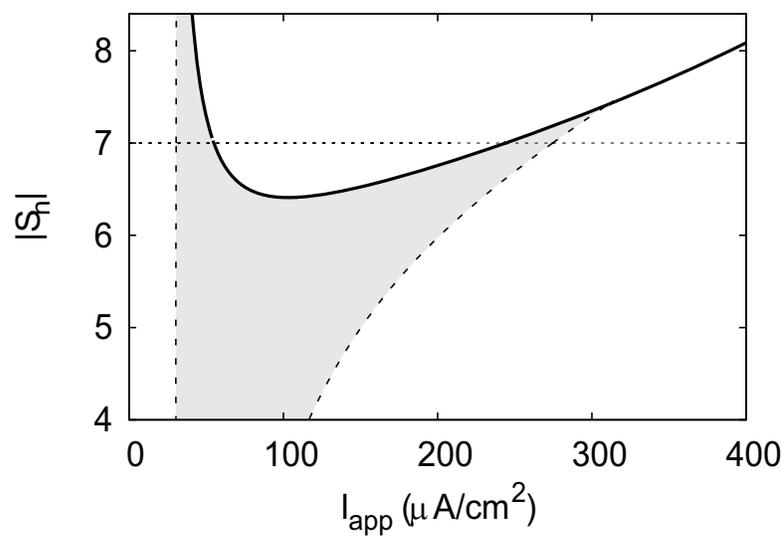
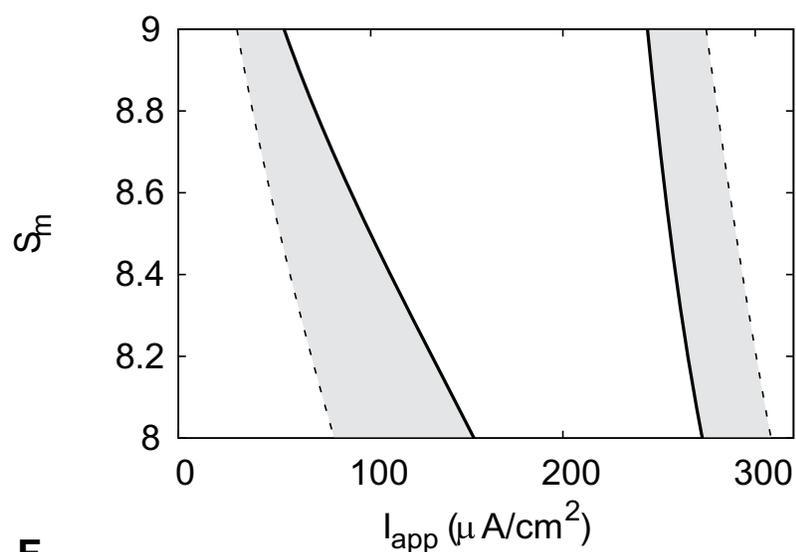
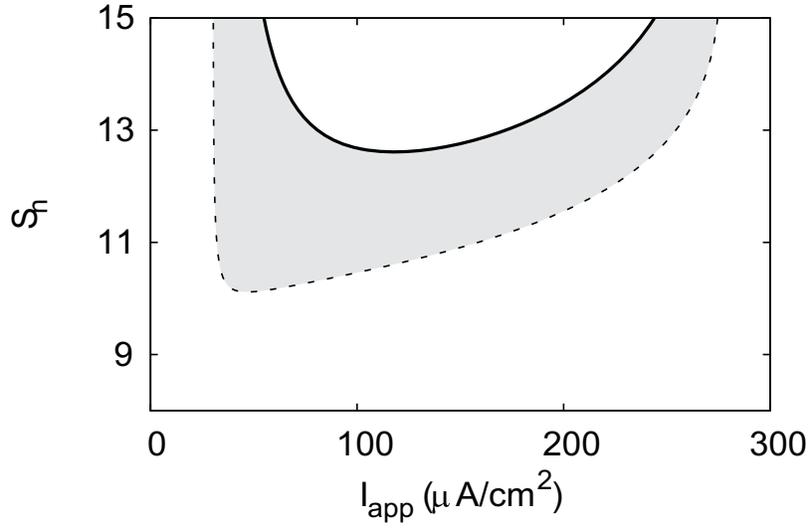

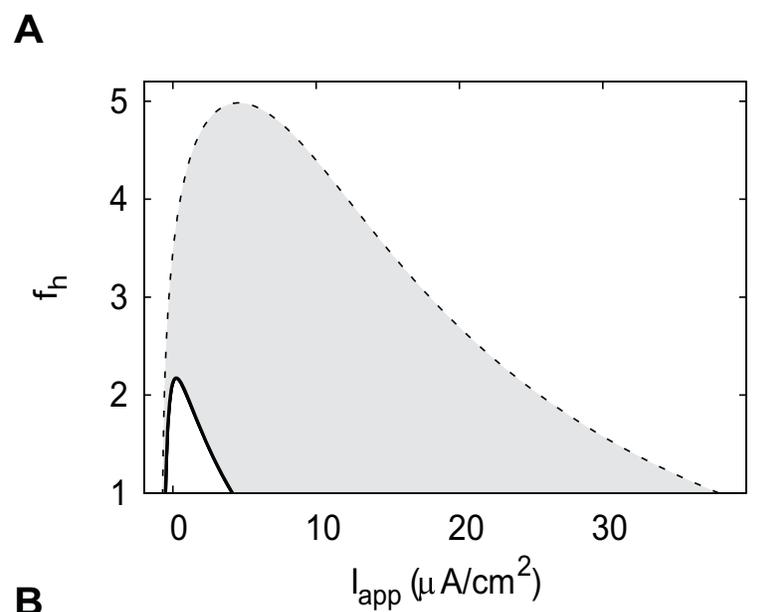
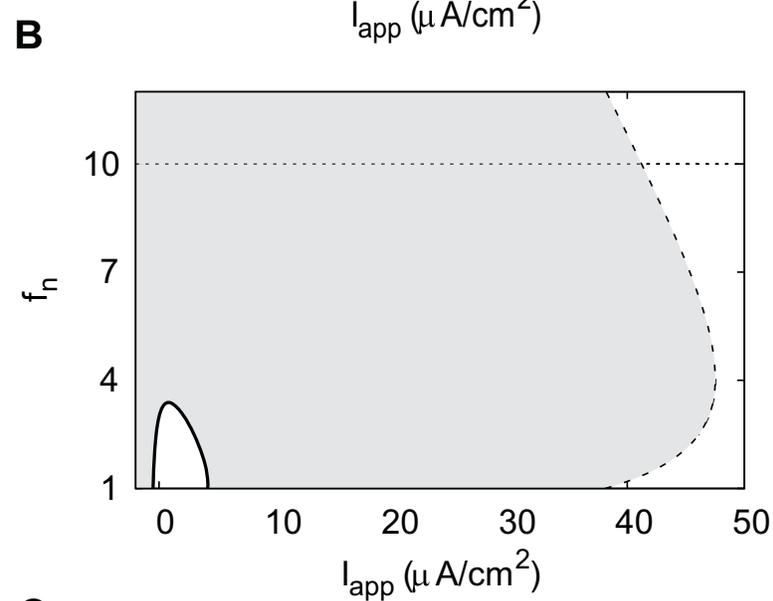
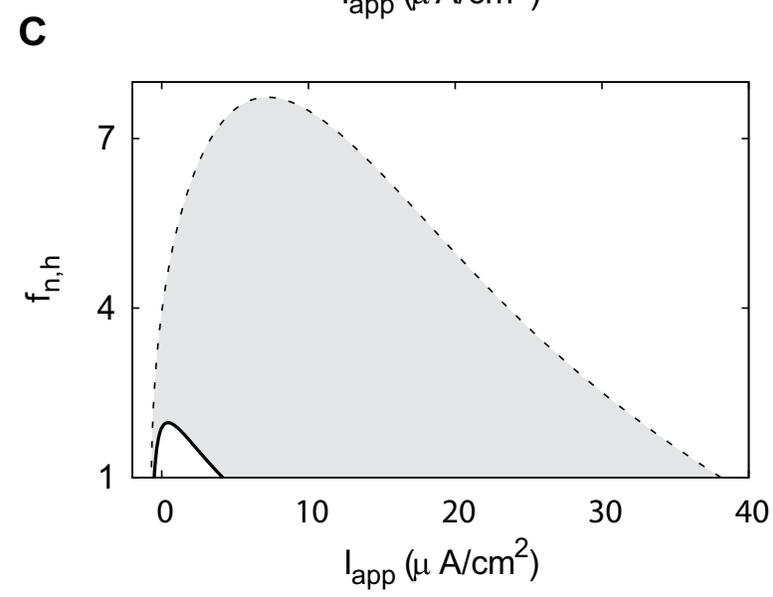
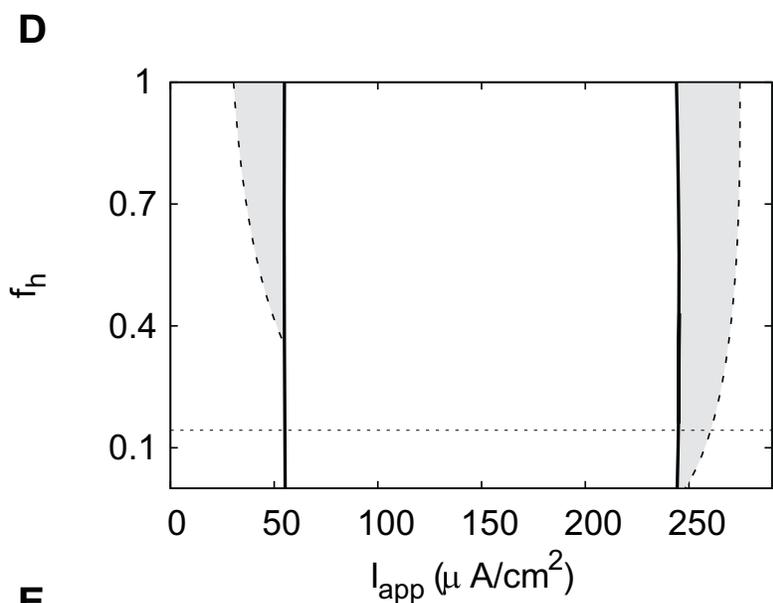
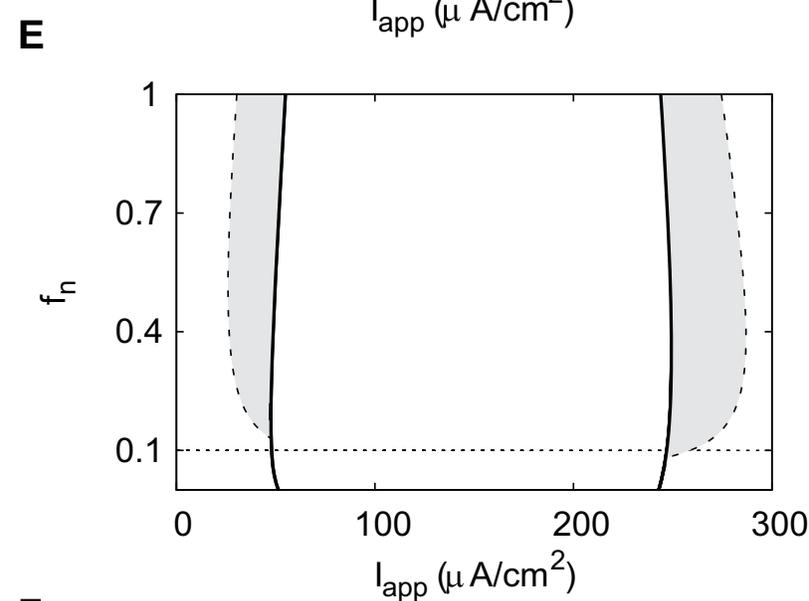
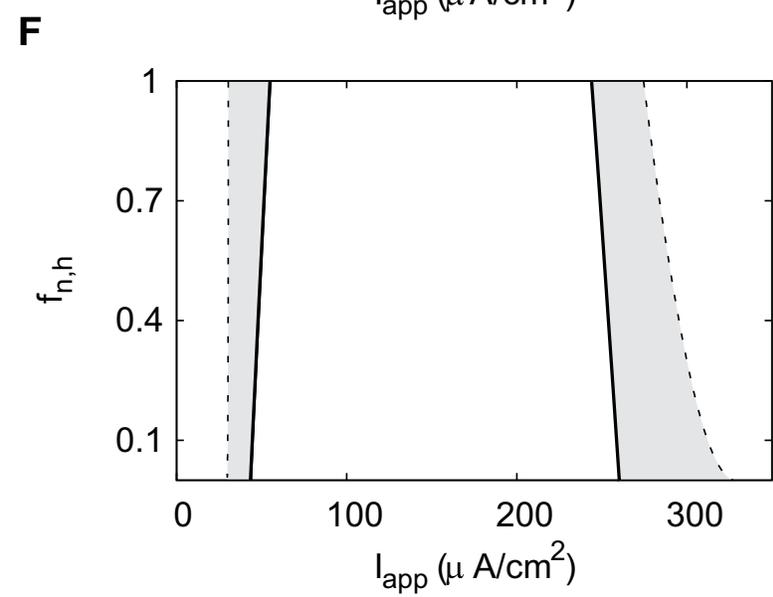

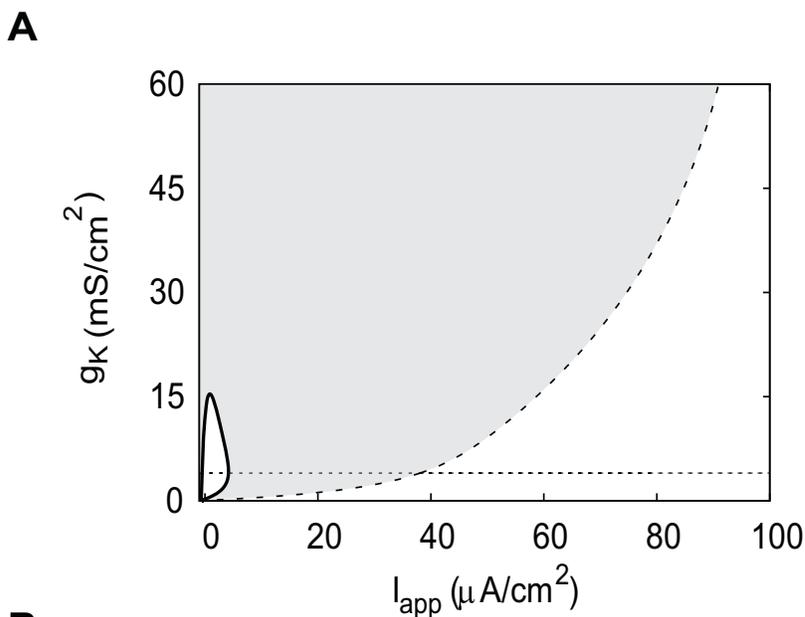
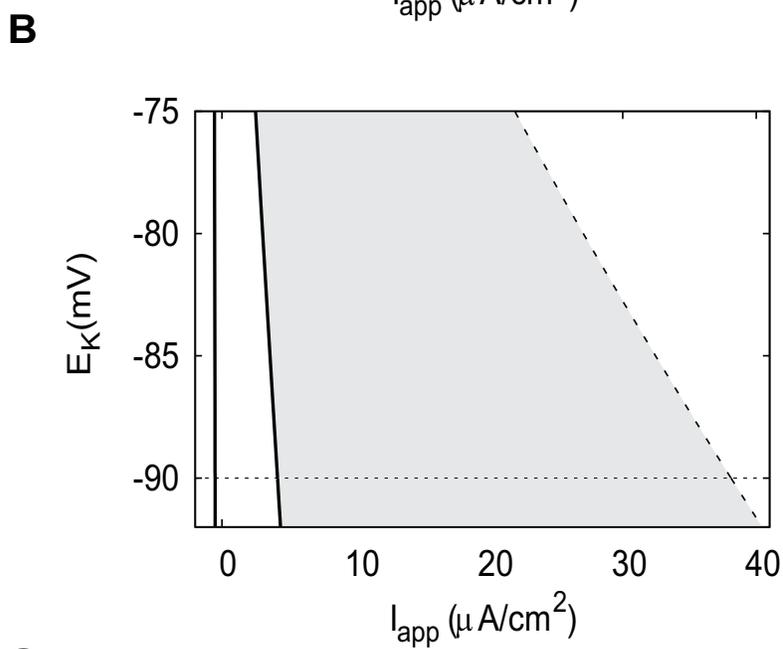
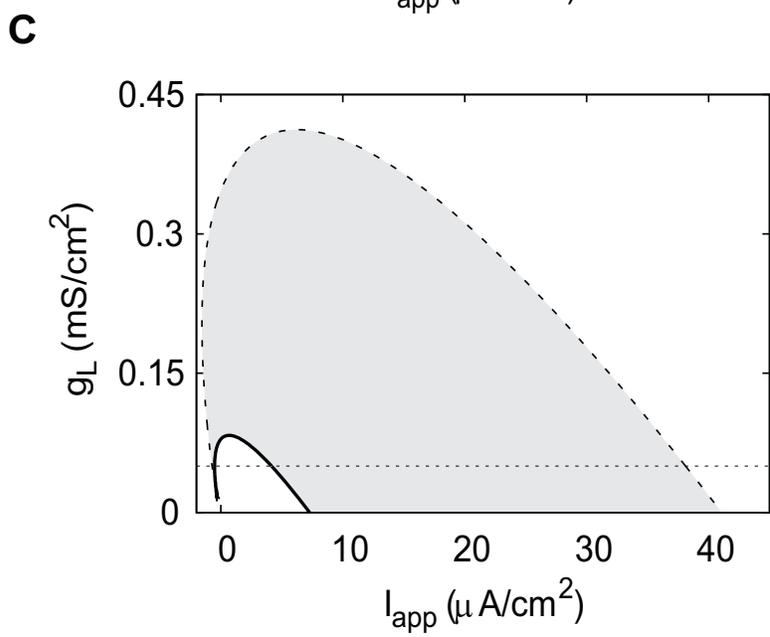
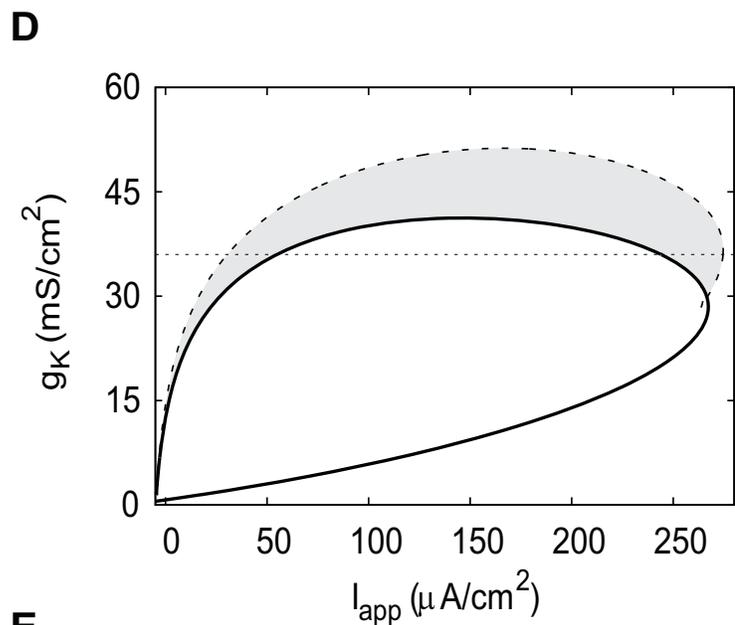
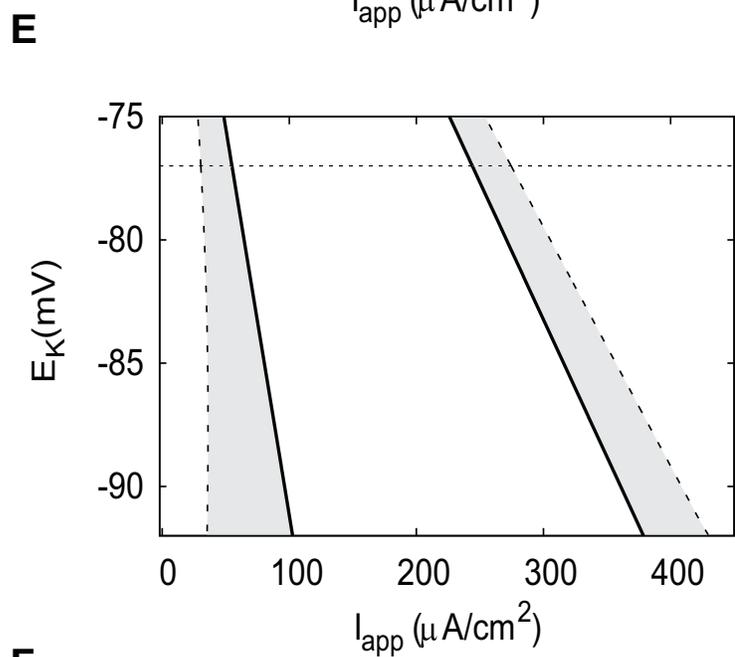
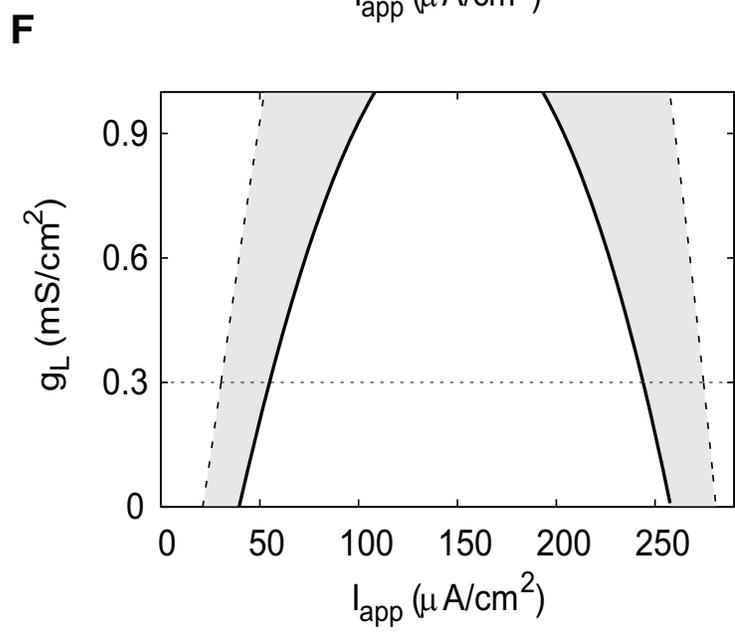